\documentclass[12pt]{article}

\usepackage{amsmath,amssymb,amscd,subfigure}
\usepackage{url}
\usepackage{graphicx} 
\usepackage{bmpsize} 
\usepackage{tabularx}
\usepackage{booktabs}
\usepackage{bm}

\def\beq{\begin{equation}}
\def\eeq{\end{equation}}
\def\barr{\begin{eqnarray}}
\def\earr{\end{eqnarray}}

\begin{document}

\begin{titlepage}

\begin{center}
\hfill  \quad  \\
\vskip 0.5 cm
{\Large \bf Anisotropic Friedel oscillations and isoperimetric transport in two-dimensional free lattice Fermi seas}

\vspace{0.5cm}

Guillermo~R.~Zemba\\
\medskip
{\em Facultad de Ingenier\'ia y Ciencias Agrarias,  Pontificia Universidad Cat\'olica Argentina,}\\
{\em  Av. Alicia Moreau de Justo 1500,(C1107AAZ) Buenos Aires, Argentina}\\
\medskip
{\em  and}\\
\medskip
{\em Departamento de F\'{\i}sica Te\'orica de Interacciones Fundamentales y Sistemas Complejos, Laboratorio Tandar,}\\
{\em  Comisi\'on Nacional de Energ\'{\i}a At\'omica,} \\
{\em Av.Libertador 8250,(C1429BNP) Buenos Aires, Argentina}\\
\end{center}
\vspace{.3cm}
\begin{abstract}
\noindent
Two physical consequences of the recent topological classification of 2d Fermi seas for free particles based on flat orbifolds $\mathbb{R}^2/\Gamma$ of discrete symmetry $\Gamma$, are explored.
The first is a study of the spatial density behavior inside the mesoscopic window enclosed by the Fermi and 
(infinite) thermal lengths.
It is shown that the single-particle density matrix exhibits a sharp directional anisotropy, encoding a real space diffraction pattern, {\it i.e.}, a Fraunhofer-like diagram of the Fermi sea, that reveals the point-group symmetries of the background lattice $\Gamma$ in the form of anisotropic Friedel oscillations.
As examples, the special cases of the $p4m$ and $p6m$ crystallographic flat orbifolds are treated in some detail. 
The second physical consequence is the establishment of an explicit geometric correspondence between momentum-space packing and electrical transport. 
A mean-field expression for the longitudinal electrical conductivity $\sigma_{xx}$ as a function of the {\it isoperimetric parameter} $\cal{I}$ of the occupied orbifold domain by utilizing geometric measure theory is presented,
satisfiying the scaling law $ \sigma_{xx} \propto 1/\sqrt{\cal{I}}$.
The topological transport formula naturally incorporates the vanishing conductivity ($\sigma_{xx} \to 0$) of the previously described ``geometric insulators,'' promoting the isoperimetric parameter as an efficient descriptor for topological and geometric phases in conductors.
\end{abstract}
\vskip 0.5 cm
\end{titlepage}
\pagenumbering{arabic}
\section{Introduction} 

Fermi seas are compact objects in momentum space of characteristic size given by the Fermi momentum $p_F$,
associated to a specific physical system \cite{lanlifs,ashmer}. Their boundary, the Fermi surfaces, determine the physical properties 
of the system in coordinate space. These may be obtained by examination of the Fourier or similar transformation
of the Fermi seas, establishing a correspondence between momentum and coordinate spaces. The characteristic size domain
in which these properties become manifest is given by scales larger or of the order of the Fermi wavelength 
$\lambda_F =h/p_F$. This general program may be completed in detail for some simple systems. 
Focus is made in this paper on 2d free gases 
of spinless fermions moving on a geometric background with lattice symmetries $\Gamma$, so as to explore the consequences 
of the topological classification of Fermi seas based on flat orbifolds $\mathbb{R}^2/\Gamma$ given in \cite{zembatop}.
The Fermi surface will be referred to, therefore, as the Fermi boundary,
given its one-dimensional character. 

Because effective field theories (EFTs) \cite{polch} only address long-range or low-energy physics, separating 
scales and absorbing high-energy effects into a restricted set of parameters, they are usually unaffected by microscopic details. However, {\it discrete symmetries} of the microscopic lattices at the mesosocopic scale
may modify this general paradigm, as has been pointed out in \cite{boze}. The general results presented here 
support this viewpoint.

In Section 2 it is shown that the ground state of non-interacting 2d fermions described by crystalline flat orbifolds $\mathbb{R}^2/\Gamma$ in momentum 
space break continuous rotational symmetry, modifying the Fermi boundary as a sharp diffraction aperture that maps onto coordinate space 
as a Fraunhofer diagram. Accounting for Wyckoff multiplicities and stabilizer fractions at orbifold boundary facets and conical vertices yields 
a modified particle number relation regarding non-crystalline case. Numerical evaluation of $p4m$ and $p6m$ $\Gamma$-domains expose 
directional phase-interference cross-star patterns, vertex beats, group-velocity focusing that slows spatial decay in correlation functions from $r^{-3/2}$ to $r^{-1}$ along flat facets, and Berry curvature directing at sharp angular apices.

In Section 3, a direct geometric correspondence between momentum-space domain packing efficiency, in terms of the isoperimetric ratio $\mathcal{I}$, and longitudinal mean-field electrical conductivity $\sigma_{xx}$ for the 17 orbifold classes is given. The latter are divided into 7 boundary-less geometric insulators 
($\sigma_{xx}=0$) and 10 classes of conductors. Generalizing transport yields a Drude-Sommerfeld-like scaling $\sigma_{xx} \propto 1/\sqrt\mathcal{I}$, naturally capturing Van Hove singularity structural kinks, non-orientable topological inflation (Möbius/Klein bottle baseline shifts) and corner-scattering relaxation time degradation.

Brief conclusions are given at the end.

\section{Mesoscopic anisotropy and the Fermi sea Fraunhofer-like diagram}
\label{sec:fraunhofer}

In a quantum statistical mechanics description, a system is governed by two distinct length scales that establish different physical properties:
{\it i)} the thermal de Broglie wavelength $\lambda_{T} = h/{\sqrt{2\pi m k_B T}}$,  which measures the thermal uncertainty of a particle's position, where 
$m$ is the mass of a non-relativistic particle and $T$ the absolute temperature. It separates classical and quantum regimes in a gas. As $T \to 0$, $\lambda_{T} \to \infty$, looses its resolution power.
{\it ii)} the Fermi wavelength $\lambda_F = {2\pi}/{k_F}$, ($k_F = p_F/\hbar$) which is the spatial scale of the highest-energy occupied state in the Fermi sea and depends solely on particle density $n$ (in 2d, $n = g_s k_F^2 / 4\pi$, where $g_s =2s+1$ is the spin degeneracy factor). It defines the average inter-particle spacing ($d \approx \lambda_F$) and represents the size of the "quantum box" or spatial footprint that each fermion carves out for itself via Pauli exclusion.
The results presented in this work apply to a fermion gas in the mesoscopic regime, that is, for distances $r$ satisfying $\lambda_F \ll r \ll \lambda_T$. 

\subsection{Isotropic free Fermi sea}
At zero temperature, an ideal non-interacting 2d Fermi gas is completely characterized in momentum space by 
its Fermi sea $\Omega_F$, which is defined by the region $|\mathbf{p} | \leq p_F$, where $p_F$ is the Fermi momentum.
In terms of the wavevector $\mathbf{k} = \mathbf{p}/\hbar$, the distribution defines a circular disk of radius $k_F$.
The coordinate space behavior of $\Omega_F$ is determined through the equal-time single-particle density matrix (or Green's)
function, {\it i.e.}, the two-point correlation function of non-relativistic free fermion fields $\psi(\mathbf{r})$:
\begin{equation}
\rho(\mathbf{r},{0})\ =\ \langle \Omega | \psi^\dagger(\mathbf{r}) \psi(\mathbf{0}) | \Omega \rangle\ ,
\end{equation}
where 
\begin{equation}
\psi(\mathbf{r}) = \int \frac{d^2\mathbf{k}}{(2\pi)^2} \, e^{i\mathbf{k}\cdot\mathbf{r}} c_{\mathbf{k}}, \quad \psi^\dagger(\mathbf{r}) = \int \frac{d^2\mathbf{k}}{(2\pi)^2} \, e^{-i\mathbf{k}\cdot\mathbf{r}} c_{\mathbf{k}}^\dagger\ ,
\end{equation}
and $c_{\mathbf{k}}$ and $c_{\mathbf{k}}^\dagger$ are fermionic Fock space operators
satisfying the usual anticommutation relations. 
Evaluating the equal-time correlation function in the Fermi sea ground state $|\Omega \rangle = \prod_{k \le k_F} c_{\mathbf{k}}^\dagger |0\rangle$,
where $| 0\rangle$ is the Fock vacuum, yields:
\begin{equation}
\langle \Omega | \psi^\dagger(\mathbf{r}) \psi(\mathbf{0}) | \Omega \rangle = \int \frac{d^2\mathbf{k}_1}{(2\pi)^2} \frac{d^2\mathbf{k}_2}{(2\pi)^2} e^{-i\mathbf{k}_1\cdot\mathbf{r}} \langle \Omega | c_{\mathbf{k}_1}^\dagger c_{\mathbf{k}_2} | \Omega \rangle\ ,
\end{equation}
with ground-state occupation $\langle \Omega | c_{\mathbf{k}_1}^\dagger c_{\mathbf{k}_2} | \Omega \rangle = (2\pi)^2 \delta^{(2)}(\mathbf{k}_1 - \mathbf{k}_2) \theta(k_F - k_1)$. Therefore 
\begin{equation}
\langle \psi^\dagger(\mathbf{r}) \psi(\mathbf{0}) \rangle = \int \frac{d^2\mathbf{k}}{(2\pi)^2} \, \theta(k_F - k) e^{-i\mathbf{k}\cdot\mathbf{r}}\ .
\end{equation}
Evaluation yields the spatial correlation amplitude profile:
\begin{equation}
\rho(\mathbf{r},{0}) = \frac{k_F}{2\pi  r} J_1\left(k_F r\right)\ .
\label{rhofree}
\end{equation}
We note that $\rho(\mathbf{r},{0})$ is not a physical local particle density. 
Instead, it defines the off-diagonal single-particle density matrix element for a 
translationally invariant system:
\begin{equation}
\rho(\mathbf{r}_1,\mathbf{r}_2) =  \rho(\mathbf{r}_1-\mathbf{r}_2,{0}) \equiv \rho(\mathbf{r},{0})\ .
\end{equation}
The quantity $\rho(\mathbf{r},{0})$ embodies two specific physical mechanisms: {\it i) Quantum non-locality}, which describes the quantum mechanical amplitude that removing (annihilating) a fermion at the origin ($0$) instantaneously creates a state that phase-locks with a fermion located at position $\mathbf{r}$.
{\it ii) Statistical phase matrix}, which evaluates the geometric phase relationship between matter-waves at two separate spatial coordinates:
positive regions ($\rho(\mathbf{r},{0}) > 0$) were waves are in-phase, leading to constructive quantum interference, and 
negative regions ($\rho(\mathbf{r},{0}) < 0$) in which waves are entirely out-of-phase, establishing destructive quantum interference.

For a perfect, homogeneous, and unperturbed 2d Fermi gas, evaluating the real observable particle density operator gives a 
spatially constant and strictly positive value: 
\begin{equation}
n(\mathbf{r})\ \equiv \rho(\mathbf{r},\mathbf{r}) = \rho(0,0)= \frac{k_F^2}{4\pi} = n_0\ .
\end{equation}
No spatial oscillations exist in the density profile of a clean, translationally invariant ground state.
However, the alternating phase profile of $\rho(\mathbf{r},{0})$ governs the structure of the multi-particle state via the pair correlation function $g(\mathbf{r})$, which determines the probability of finding two fermions in the same state ({\it e.g.},with parallel spins) simultaneously separated by a spatial displacement $\mathbf{r}$:
\begin{equation}
g(\mathbf{r}) = 1 - \left( \frac{\rho(\mathbf{r},{0}))}{\rho({0},{0})} \right)^2 = 1 - \left[ \frac{2 J_1(k_F r)}{k_F r} \right]^2\ .
\end{equation}
As $r \to 0$, the ratio approaches unity, rendering $g(0) =0$ which is a consequence of the Pauli exclusion principle. 
For $r > 0$, $g(\mathbf{r})$ smoothly oscillates as it asymptotically reaches 1. This means every fermion carries a localized, dynamically breathing clearance bubble known as a Pauli hole, forcing neighboring identical particles into concentric coordinate shells.

To map these hidden ground-state correlations onto a real, observable, spatially oscillating change in local particle density ($\delta n(\mathbf{r})$), continuous translational symmetry must be broken. Introducing a physical constraint—such as a localized potential defect, a step edge, or an artificial light barrier—forces incoming fermion waves to back-scatter.
The resulting interference between incoming and reflected trajectories manifests as physical standing waves. This local spatial density modulation maps to the squared modulus of the correlation amplitude:
\begin{equation}
\delta n(\mathbf{r}) \propto |\rho(\mathbf{r},{0})|^2 \propto \frac{J_1^2(k_F r)}{r^2}\ .
\end{equation}
For large $r$, it behaves as:
\begin{equation}
\delta n(\mathbf{r}) \propto \frac{\cos^2\left(k_F r - \frac{3\pi}{4}\right)}{r^3}  \propto \frac{1 + \sin(2k_F r)}{r^3}\ .
\end{equation}
The physical consequences of this are two: firstly, the physical density ripples oscillate at twice the Fermi wavevector ($2k_F$), which stems from momentum conservation: a forward-moving state at $+k_F$ back-scatters off the obstacle into a backward-moving state at $-k_F$, with net momentum transfer $2k_F$.
Secondly, in 2d the amplitude of these density variations decays as $1/r^3$. This attenuation represents spatial quantum screening, ensuring that localized boundaries or defects do not globally destabilize the Fermi sea.
Known as Friedel oscillations, these density fluctuations alternate between high and low density, but the absolute total local density ($n_0 + 
\delta n(\mathbf{r})$) remains strictly positive \cite{friedel}.

While isolated chemical defects spark Friedel oscillations, non-impurity constraints that break continuous translational symmetry can also reveal them.
For example, in crystal step edges on clean single-crystal surfaces (e.g., $\text{Cu}(111)$ or $\text{Ag}(111)$) which act as planar barriers such that bouncing electrons interfere with incoming waves, setting up a 1d standing wave pattern measured via Scanning Tunneling Microscopy (STM) \cite{jean}.
Quantum Corrals give another example: as shown in relevant IBM experiments, positioning substrate atoms in a closed circle creates a boundary that confines the 2d electron gas inside. The interior oscillations are pure standing wave eigenstates of a clean, bounded system \cite{ibmexp}.
Finally, synthetic ultracold gases: neutral fermionic atoms (e.g., $^{6}\text{Li}$, $^{40}\text{K}$) trapped by lasers in flat 2d sheets contain zero lattice defects. Abruptly introducing a sharp wall of light forces back-scattering, which can be directly recorded via in-situ optical absorption imaging
\cite{edge,cheuk,cheng,hueck,krinner,reinaudi}.

\subsection{Anisotropic Fermi sea orbifold domains}

For a non-interacting fermion gas moving on a crystallographic background the previous analysis is modified \cite{zembatop},
since the Fermi sea now becomes a flat orbifold ${\cal O}= \mathbb{R}^2/\Gamma$, where $\Gamma$ is one of the discrete space groups (wallpaper groups). See table (\ref{tab:zemba_classes}) for a quick remainder.
Unlike a continuum electron gas where the single-particle density matrix 
decays isotropically via Friedel oscillations, the presence of the background lattice crystallographic symmetry $\Gamma$ induces a breaking of 
continuous rotational symmetry, leading to anisotropy.
The computation of $\rho(\mathbf{r}, \mathbf{r'}) $ is done by performing the partial Fourier transform over the compact momentum domain 
defined by the flat orbifold. Because $T=0$, the distribution function remains a strict Heaviside step function, causing the momentum-space 
boundary to act as a sharp aperture.
This framework leads to the Fraunhofer diffraction analogy:
the local density of states (LDOS) and $\rho(\mathbf{r}, \mathbf{r'}) $ take the mathematical form of a spatial diffraction pattern, where 
the sharp, non-spherical Fermi surface acts exactly as a diffraction aperture. 
The resulting real-space pattern yields a directional map of lobes and phase-shifted oscillations corresponding to the Fraunhofer-like diagram of the Fermi sea.
This diagram maps out the corners, flat facets, and local curvature of the underlying orbifold directly onto real-space mesoscopic scales.

The total particle number $N$ for a 2d spinless (or polarized) non-interacting fermion gas in a real-space area $A$ is given by integrating the equal-time single-particle density matrix $\langle \Omega | \psi^\dagger(\mathbf{r})\psi(\mathbf{r}) | \Omega \rangle$ over coordinate space:
\begin{equation}
N = \int d^2r \, \langle\Omega|\psi^\dagger(\mathbf{r})\psi(\mathbf{r})|\Omega\rangle = \frac{A}{(2\pi\hbar)^2} \int_{\Omega_F} d^2p\ ,
\label{naniso}
\end{equation}
where $d^2p/(2\pi\hbar)^2$ is the standard phase-space density of states per area $A$ and $\Omega_F$ is the Fermi sea. 
For simplicity, spin and valley degeneracies are ignored in this analysis.

The case with $\Omega_F=\mathbb{R}^2$ was discussed previously. For an orbifold of the type $\Omega_F={\cal O}= \mathbb{R}^2/\Gamma$, 
the integration over ${\cal O}$  may be performed with the Satake orbifold integration theory.
Schematically, it decomposes the regions of integration in a single, bounded fundamental 
domain $\Omega_\Gamma$ plus boundary contributions:
\begin{equation}
\int_{\cal O} d^2p = \int_{\mathbb{R}^2/\Gamma} d^2p = \text{Area}(\Omega_\Gamma \cap \text{Fermi boundary})\ .
\label{normaliz}
\end{equation}
The measure decomposition is based on Ichiro Satake’s foundational 1956–1957 theory of V-manifolds \cite{satake}, now called orbifolds, and provides a 
framework for integrating functions, particle densities, or differential forms over a quotient space $\mathcal{O} = X/\Gamma$ by stratifying 
the space into its regular and singular geometric strata (interior bulk, boundary edges, and point singularities) and weighting each stratum 
by the inverse order of its local isotropy (stabilizer) group.
Eq. (\ref{naniso}) becomes:
\begin{equation}
N = \frac{A}{(2\pi\hbar)^2} \left[ \int_{\cal O} d^2p + \frac{1}{2}\int_{\partial\cal O} d^2p + \sum_{v} w_v \right]\ .
\label{Nfororb}
\end{equation}
The first term on the r.h.s. is the bulk stratum. Interior points of the irreducible fundamental domain $\Omega_\Gamma$ have trivial generic stabilizers ($\vert{}G\vert{}=1$) and have integration weight equal to 1. 
The second term is the edge stratum: 1d boundary edges ($\partial\mathcal{O}$) correspond to mirror (reflection) planes in the wallpaper group $\Gamma$. Each reflection wall is shared symmetrically by 2 adjacent tiling wedges in the universal 
cover $\mathbb{R}^2$ and the contribution is $1/2$. 
The third term is the vertex (conical singularity) stratum: isolated corners ($v = \Gamma, X, M, K$) are fixed points of point-subgroups (cone singularities with opening angles like $\pi/4, \pi/3, \pi/2$). Their contribution has weight  $w_v = {1}/{\vert{}G_v\vert{}}$ or the Wyckoff site multiplicity fraction corresponding to the point-group order of the local stabilizer ({\it e.g.}, $w_\Gamma = 1/8$ for $p4m$, $1/12$ for $p6m$), ensuring that multi-fold orbit intersections don't over-count or under-count the total quantum state phase-space volume $N$ \cite{crystal,chadi}.

From the previous discussion on the normalization of the single-particle density matrix 
$\rho(\mathbf{r},{0})$ it becomes clear that obtaining an analytic expression for (\ref{rhofree}) becomes a demanding and unnecessary task.
Among the factors that challenge this idea are: 

{\it i) Polygon-wedge aperture diffraction:} in momentum space, restricting the integral 
to an orbifold fundamental domain intersecting a Fermi boundary (which warps or squares off with filling) 
is equivalent to computing the Fourier 
transform of a non-circular, sharp-cornered polygonal (conical) sector. In diffraction theory, Fourier transforms of polygonal apertures can be 
written analytically using edge-wave or vertex-diffraction contour formulas ({\it e.g.}, Rubinowicz-type boundary diffraction waves).
However, these 
yield sums of directional phase-shifted Fresnel(sinc)-like edge contributions whose orientation depends explicitly on the Wyckoff boundary angles 
($\pi/4, \pi/3, \pi/2$) and the local normal vectors of the warped Fermi surface. Writing this out yields a multi-term geometric bookkeeping 
large structure rather than a transparent scaling formula.

{\it ii) Filling-fraction dynamic shape:} unlike a circle (parameterized solely by $k_F$), an orbifold 
Fermi boundary changes topological shape and hits Van Hove saddle points as filling varies. An analytic expression for $\rho(\mathbf{r},{0})$ would have to be piecewise-defined across Lifshitz transition boundaries. 

On the one hand, Eq. (\ref{Nfororb}) handles phase-space counting rigorously via Satake-Wyckoff boundary and vertex weights, solving the particle-number bookkeeping analytically without needing position-space coordinates. 
On the other hand, rather than a complete position-space formula for $\rho(\mathbf{r},{0})$, analytic asymptotic scaling formulas of it for the directional group velocity focusing, may be derived. In particular, it may be noted that flat facet normals slow down spatial decay along high-symmetry axes 
(shifting power laws from isotropic baselines to $r^{-1}$ along flat facets). This analysis is presented in Section 2.5. 

An alternative and efficient way of computing $\rho(\mathbf{r},{0})$ may be performed by a direct numerical sampling (Fast Fourier Transform) 
of the characteristic function on the orbifold fundamental domain ({\it e.g.}, $p4m$ 4-fold cross, $p6m$ 6-pointed star, $p2$ elliptic stretching).
This method method provides the visualization of the main characteristics of $\rho(\mathbf{r},{0})$, like symmetry-locking and vertex beats 
of the Friedel density oscillations. In the following, it is used to provide two specific examples of the above calculation.

\subsection{The square lattice orbifold ($p4m$ / $*442$)}
The space group $p4m$ combines square lattice translations with a point group symmetry of $C_{4v}$ ($4mm$), which has a group order of $|\mathcal{P}| = 8$.
The main characteristics are:
{\it i) Fundamental domain geometry:} the orbifold $\mathbb{R}^2/p4m $ restricts the independent momentum states to a 
     a $\pi/4- \pi/4 - \pi/2$ right isosceles triangle within the irreducible Brillouin zone, with vertices at $\Gamma = (0,0)$, $X = (\pi/a, 0)$, and 
     $M = (\pi/a, \pi/a)$ ($a$ is the square length). 
     {\it ii) Orbifold singularities:} the boundaries act as mirror reflection lines. The corners $\Gamma$ and $M$ form conical singularities with a local opening angle of $\pi/2$, whereas $X$ presents a sharper $\pi/4$ opening angle.
{\it iii) Statistical weighting:} a state positioned precisely at the $\Gamma$ vertex is shared among 8 symmetric replicas, yielding a vertex weight of $w_\Gamma = 1/8$ \cite{pretti}.

\begin{figure}[htbp]
    \centering 
    \includegraphics[width=1.0\textwidth]{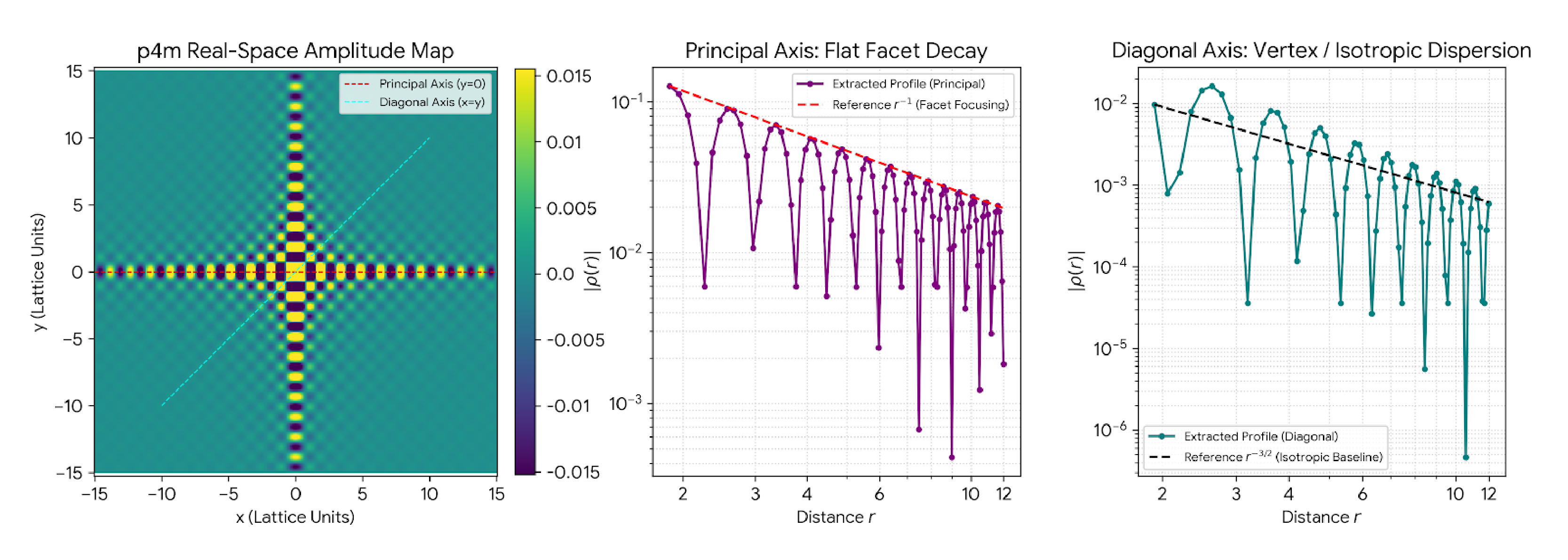} 
    \caption{$\rho(\mathbf{r},{0})$ for the $p4m$ wallpaper group displaying the Fraunhofer diagram in coordinate space (left). 
    $\rho(r)=|\rho(\mathbf{r},{0})|$ vs. $r=|\mathbf{r}|$ and anchor-normalized power-law slow reference $r^{-1}$ decay rate 
    for principal/flat-facet cut (center).  
    $\rho(r)=|\rho(\mathbf{r},{0})|$ vs. $r=\mathbf{r}$ and anchor-normalized power-law standard reference $r^{-3/2}$ decay rate 
    along diagonal vertex-crossing directions (right).}
    \label{fig:p4m} 
\end{figure}

The four-fold cross diffraction spot may be seen in figure (\ref{fig:p4m}). Two comments are in order:
{\it Friedel oscillation directionality:} density screening profiles lose circular symmetry and radiate along a four-fold ($C_4$) cross-like pattern matching the fundamental axes.
{\it Vertex beats:} the sharp $\pi/2$ and $\pi/4$ boundary intersections create step-discontinuities in angular momentum variables. Under Fourier transformation, these act as vertex contributions that generate unique spatial phase interference beats along the $\langle 11 \rangle$ diagonals.
\subsection{The triangular lattice orbifold ($p6m$ / $*632$)}
The space group $p6m$ combines hexagonal translations with a point group symmetry of $C_{6v}$ ($6mm$), possessing a group order of $|\mathcal{P}| = 12$.
The key features are:
{\it i) Fundamental domain geometry:} the orbifold shrinks the independent momentum zone into a $\pi/6 - \pi/3 - \pi/2$ scalene triangle. The vertices are designated as $\Gamma = (0,0)$, $M = (\pi/a, \pi/(\sqrt{3}a))$, and $K = (\pi/(3a), 0)$.
{\it ii) Orbifold singularities:} the geometric constraints are more stringent. The $\Gamma$ point features a $\pi/3$ angular sector, the $K$ point features a tight $\pi/6$ sector, and the $M$ point forms a right-angle $\pi/2$ intersection.
{\it iii) Statistical weighting:} A state exactly occupying the $\Gamma$ point is shared among 12 symmetric replicas, generating a small vertex weight of $w_\Gamma = 1/12$.

\begin{figure}[htbp]
    \centering 
    \includegraphics[width=1.1\textwidth]{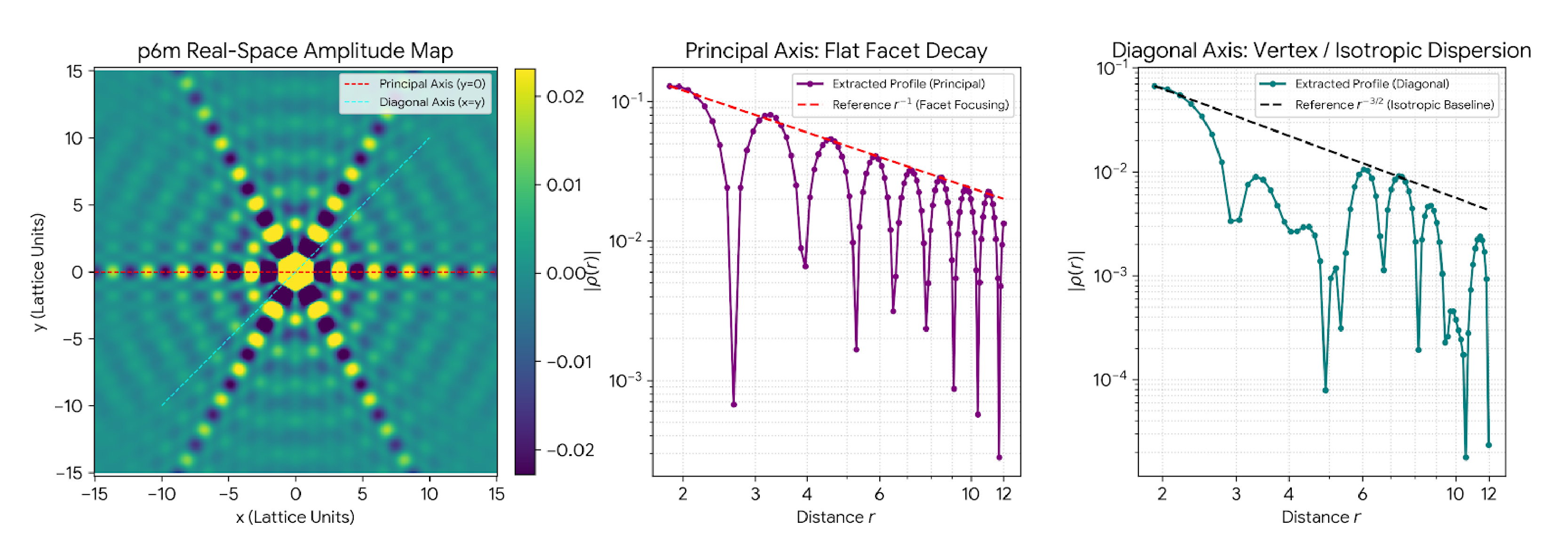} 
 \caption{$\rho(\mathbf{r},{0})$ for the $p6m$ wallpaper group displaying the Fraunhofer diagram in coordinate space (left). 
    $\rho(r)=|\rho(\mathbf{r},{0})|$ vs. $r=|\mathbf{r}|$ and anchor-normalized power-law slow reference $r^{-1}$ decay rate 
    for principal/flat-facet cut (center).  
    $\rho(r)=|\rho(\mathbf{r},{0})|$ vs. $r=\mathbf{r}$ and anchor-normalized power-law standard reference $r^{-3/2}$ decay rate 
    along diagonal vertex-crossing directions (right).}   
    \label{fig:p6m} 
    \end{figure}

A six-pointed diffraction star may be seen in figure (\ref{fig:p6m}). Two observations are in order:
{\it Hexagonal interference grids:} the real-space density matrix constructs a rigid six-fold ($C_6$) grid-like radiation pattern.
{\it Multi-frequency nesting vectors:} the geometric boundary cuts yield three distinct spatial nesting frequencies ($Q_1, Q_2, Q_3$). These interfere constructively to generate a persistent triangular lattice of charge density ripples.

\subsection{Physical consequences: charge routing and Berry curvature}
As discussed above, the boundaries of the fundamental domains enforce strict selection rules: the Fermi boundary must intersect mirror lines at exactly $\pi/2$ to remain differentiable across symmetric reflections. As the density increases, a circular Fermi sea squares off or hexagons out to meet these conditions.
Because the electron group velocity ($\mathbf{v}_g = (1/\hbar)\nabla_{\mathbf{k}}\epsilon(\mathbf{k})$, where $\epsilon(\mathbf{k})$ is the energy) is normal to the Fermi boundary, flat regions channel electrons preferentially along high-symmetry axes. This electron focusing effect alters the spatial decay power from the isotropic $r^{-3/2}$ down to a slower $r^{-1}$ along the flat facets. 
This behavior may be understood by a diffraction analogy where sharp edges become sources:
a sharp, flat-edged boundary facet in reciprocal space acts as a linear coherent aperture.
Fourier transformation of a sharp step bounded by a flat momentum-space segment produces directed wave-packet "beaming" normal to the facet.
Just as line or planar-front contributions in wave propagation attenuate more slowly geometrically than point-source spherical decay, 
the coherent phase integration along a flat reciprocal-space facet constrains transverse spreading. This shifts the asymptotic amplitude 
envelope from higher-order multidirectional interference dilution ($r^{-3/2}$) to the slower $r^{-1}$ planar/linear diffraction envelope.
Furthermore, if inversion or time-reversal symmetry is lifted, the Berry curvature accumulates at the sharp triangular vertices ($\Gamma, X, M, K$), 
turning these orbifold corners into topological conduits that shape non-local quantum transport behavior.

The real-space patterns emerging from a topologically constrained Fermi gas are mathematically isomorphic to the far field (Fraunhofer) optical diffraction profiles produced by scattering light through apertures of matching symmetry. 
The structural difference lies in the nature of the measurement: classical optical diffraction typically registers intensity of light, whereas the 
Friedel density oscillations are related to the electric field, 
preserving the alternating signs (crests and valleys) of the quantum mechanical wave amplitude profile.

\section{Isoperimetric conductivity and Fermi sea packing}
\label{sec:transport}
\subsection{Isoperimetric ratio and Fermi seas and their boundaries}
Electric transport may be accounted for by the geometric packing efficiency of the occupied momentum states, 
establishing a direct link between the geometric measure theory discussed above and electrical conductivity ($\sigma_{xx}$), 
within the framework of the orbifold classes.

In d-dimensional space, the {\it modified isoperimetric ratio} $\mathcal{I}_{\Omega}$ for a domain $\Omega$ is defined by:
\begin{equation}
    \mathcal{I}_{\Omega} = \frac{\text{Vol}(\Omega)}{\left[\text{Area}(\partial\Omega)\right]^{d/(d-1)}}\ .
    \label{isoind}
\end{equation}
In particular, $\Omega$ may represent the occupied Fermi sea,
where the volume $\text{Vol}(\Omega)$ is given by the total carrier density $n$, and $\text{Area}(\partial\Omega)$ measures the total area of the scattering boundary (the Fermi boundary).
In flat 2d geometry, the relationship between the perimeter length ($\ell$) of a simple closed curve and its enclosed area ($\cal{A}$) is fundamentally governed by the isoperimetric inequality:
\begin{equation}
    {\ell}^2 \geq 4\pi {\cal A}\ .
\end{equation}
For the specific case of 2d, the {\it isoperimetric ratio} $\mathcal{I}$ (often denoted as ${\cal Q}$ is the mathematical literature) is defined to 
quantify the efficiency of the enclosure of an area (or its geometric departure from circularity) and includes a normalization factor of $4\pi$ 
with respect to Eq. (\ref{isoind}) for convenience: 
\begin{equation}
    \mathcal{I} = \frac{4\pi {\cal A}}{{\ell}^2}, \quad \text{where } 0 < \mathcal{I} \leq 1\ .
    \label{iso2d}
\end{equation}
The upper bound $\mathcal{I} = 1$ is satisfied uniquely by a perfect circle. In a 2d quantum electron gas at temperature $T=0$, the Fermi sea acts as a frozen, sharp geometric object in momentum space. Here, the enclosed area $\cal{A}$ corresponds to the carrier density via Luttinger's theorem, while the boundary perimeter $\ell$ defines the actual Fermi boundary av s the compactness parameter) characterizes the geometric efficiency of a Fermi sea.
It is worth remarking that the upper bound in Eq. (\ref{iso2d}) excludes embeddings from higher dimensions. 

The physical insight to consider $\mathcal{I}$ is to obtain a quantitative geometric measure of 
electric conductivity, a property related to the boundaries of the Fermi seas. $\mathcal{I}$ 
compares two surfaces with the same area and distinguishes that with larger boundary length as the 
one with better conduction properties, minimizing $\mathcal{I}$. For example, for a regular $N$ polygon, 
\begin{equation}
    \mathcal{I} = \frac{\pi}{N \tan(\pi/N)}\ ,
    \label{polygn}
\end{equation}
which is minimal for $N=3$, with an approximated values of 0.605.

For Fermi surfaces in 3d (\ref{isoind}) could serve as a useful parameter for classifying Fermi seas.

\subsection{Topological classes of Fermi seas (Orbifold formulation)}

According to the topological classification for spinless (that is, polarized, with $g_s =1$) fermions \cite{zembatop}, a 2d lattice with a specific crystallographic space group induces a flat geometric orbifold ($\mathbb{R}^2/\Gamma$)
in momentum space. There are 17 distinct topological categories, establishing a binary criterion for electrical conduction based on the presence of a boundary: 

{\it i) Geometric insulators (7 Classes):} spaces without free boundaries ({\it e.g.}, matching the topology of a $2$-sphere, $2$-torus, Klein bottle, or real projective plane). Because the boundary vanishes (${\ell}=0$), there are no accessible low-energy electronic excitations at $T=0$, yielding $\sigma = 0$. However, all of these boundary-less surfaces are embeddings from 3d. For the geometric insulators, the range of $\mathcal{I}$ may be formally extended to include
the isolated value $\infty$ since it corresponds to the case $\ell = 0$.

{\it ii) Geometric conductors (10 Classes):} spaces possessing open boundaries ({\it e.g.}, disk topologies, annuli, and M{\"o}bius strips). The boundary hosts gapless modes capable of hosting net transport under an applied external field.

This characterization was given in \cite{zembatop} and it provides a geometric or kinematical definition of 
charge transport, avoiding the more detailed dynamics aspects.
The boundary perimeter $\ell$ of the Fermi surface and the enclosed area $\cal{A}$ of the Fermi sea are strictly regulated by the topological boundaries of these groups.
It is mathematically impossible for the isoperimetric parameter to remain constant across the 17 
distinct topological classes due to three fundamental geometric conditions:

{\it i) Boundary singularities:} the conductor classes form disk topologies bounded by sharp reflection corners. 
The insulator classes fold into closed 2-sphere topologies containing discrete conical
singularities. The remaining less familiar classes form non-orientable topologies such as the M\"obius strip or Klein bottle. Each boundary type alters the metric space available for perimeter expansion.

{\it ii) Anisotropic spatial constraints:} a highly symmetric group like the hexagonal lattice ($p6m$) forces isotropic growth, keeping the baseline quotient high ($\mathcal{I} \approx 0.907$). A low-symmetry group like the rectangular lattice ($pmm$) decouples the spatial axes ($a \neq b$), permitting the Fermi sea to elongate arbitrarily. This allows 
$\mathcal{I} \to 0$ and drives the reverse isoperimetric parameter toward infinity ($\infty$).
    
{\it iii) Continuous parameter invariants:} while the topological class is a discrete invariant, the actual shape of the Fermi surface boundary evolves continuously as a function of the electron filling fraction \cite{zembatop}.

Table \ref{tab:zemba_classes} summarizes the 2d orbifold classes and includes typical values of $\mathcal{I}$.
In order to establish a more detailed correspondence, a derivation of electrical conductivity is convenient.
\begin{table}[htbp]
\centering
\caption{Isoperimetric values and quantum phases of the 17 Orbifolds}
\label{tab:zemba_classes}
\vspace{0.3cm}
\begin{tabular}{llcll}
\toprule
\textbf{$\Gamma$} & \textbf{Geometry}  & $\mathcal{I}$ \\ 
\midrule
\multicolumn{5}{l}{\textit{Conducting classes (open topologies with reflection boundaries/cuts, ${\ell} > 0$)}} \\
\midrule
\texttt{*2222} & Rectangle (Aspect ratio $r$)  & ${\pi r}/{(1+r)^2}$ (max $\approx 0.785$)  \\
\texttt{*333}  & Equilateral Triangle    & ${\pi}/{(3\sqrt{3})} \approx 0.605$ \\
\texttt{*442}  & Isosceles Right Triangle  & ${\pi}/{(2(1+\sqrt{2})^2)} \approx 0.269$  \\
\texttt{*632}  & $\pi/6$-$\pi/3$-$\pi/2$ Triangle  & ${\pi}/{(3\sqrt{3}(1+\sqrt{3})^2)} \approx 0.081$ \\
\texttt{22*}   & Rectangle with half-turns  & ${\pi w h}/{(w+h)^2}$ (max $\approx 0.785$)  \\
\texttt{3*3}   & Rhombus ($60^\circ$)  & ${\pi \sqrt{3}}/{8} \approx 0.680$ \\
\texttt{4*2}   & Square sector  & ${\pi}/{(1+\sqrt{2})^2} \approx 0.539$ \\
\texttt{6*2}   & Hexagonal sector  & ${\pi\sqrt{3}}/{9} \approx 0.605$ \\
\texttt{*o}    & Annulus (open cylinder) & $|r_2 -r_1|/(r_2 +r_1)$ \\
\texttt{x*}    & Open M\"{o}bius strip  & $\pi w/l$ ($1/2$ twist)  unbounded twist-dependent \\
\midrule
\multicolumn{5}{l}{\textit{Insulating classes (closed 3d embbedded topologies with no 1d boundaries, ${\ell}= 0$)}} \\
\midrule
\texttt{2222}  & 2-Sphere with conical points  & $\infty$ \\
\texttt{333}   & 2-Sphere with conical points &  $\infty$  \\
\texttt{442}   & 2-Sphere with conical points &  $\infty$  \\
\texttt{632}   & 2-Sphere with conical points &  $\infty$  \\
\texttt{o}     & Regular torus               &  $\infty$ \\
\texttt{xX}    & Klein bottle                &  $\infty$  \\
\texttt{q}     & Real Projective Plane       &  $\infty$ \\
\bottomrule
\end{tabular}
\end{table}

The fundamental differences between a highly symmetric constraint ($p6m$) and a highly unconstrained system ($pmm$) are summarized explicitly in Table \ref{tab:geo_comp}.
\begin{table}[h]
\centering
\caption{Geometric comparison of lattice constraints}
\label{tab:geo_comp}
\vspace{0.2cm}
\small
\begin{tabular}{lcc}
\toprule
\textbf{Geometric property} & \textbf{hexagonal lattice ($p6m$)} & \textbf{rectangular lattice ($pmm$)} \\
\midrule
Rotational symmetry & high (6-fold) & low (2-fold) \\
Deformation behavior & isotropic / compact & anisotropic / highly stretchable \\
Boundary corners & $\pi/6, \pi/3, \pi/2$ & strictly $\pi/2$ \\
$\mathcal{I}$ & $\approx 0.907$ & variable ($\to 0$) \\
\bottomrule
\end{tabular}
\end{table}
\subsection{Some physical consequences of the isoperimetric ratio and Fermi seas and their boundaries behavior}

Given the topological nature of the orbifold classification of \cite{zembatop}, it is possible to explore 
some of the physical consequences implied by the isoperimetric ratio for systems that may be considered as 
extensions of the classes considered so far.

{\it Behavior of $\mathcal{I}$ at Van Hove singularities:}
when the filling fraction of the Fermi sea is varied continuously, $1/\mathcal{I}$ encounters a non-analytical discontinuity (a structural kink) precisely at a Van Hove Singularity (VHS). This critical event marks a topological Lifshitz transition where the Fermi surface boundary touches the edge of the Brillouin zone orbifold, causing the gradient of the energy band to vanish, $\nabla_{\mathbf{k}} \epsilon_\mathbf{k} = 0$, 
where $\epsilon_\mathbf{k}(\mathbf{k})$ is the energy. 
The geometric consequences of the VHS on the isoperimetric ratio are threefold:
{\it i) Derivative discontinuity:} the parametric curve of $1/\mathcal{I}$ develops a cusp, signaling that the boundary curvature has flattened against a reflection symmetry wall.
{\it ii) Topological ruptures:} in anisotropic groups like $pmm$, the boundary undergoes a physical rupture, transitioning from a single closed disk topology into an open, periodic cylinder (neck topology). This allows the perimeter to expand rapidly relative to the interior area.
{\it iii) Density of states divergence:} because the perimeter length $\ell$ of the Fermi surface is directly proportional to the integrated density of states ($N(\epsilon_\mathbf{k})$) at that energy level:
    \begin{equation}
    N(\epsilon_\mathbf{k}) \propto \oint_{\partial {\cal O}} \frac{d\ell}{|\nabla_{\mathbf{k}} \epsilon_\mathbf{k}|}\ .
    \end{equation}
    The vanishing of the denominator causes a logarithmic or step-like divergence in the phase space. This triggers a maximum localized growth of the value of $1/\mathcal{I}$ at the point of the topological transition.
Although $1/\mathcal{I}$ controls geometrical perimeter inflation relative to enclosed area $\mathcal{A}$, 
VHS divergence affects the spectral weight/scattering matrix elements rather than breaking Luttinger-enforced 
carrier density scaling $n(\mu)$ as a function of the chemical potential $\mu$.   
    
{\it Electronic instabilities induced by the singularity:} 
the non-analytical breakpoint induced by the Lifshitz transition acts as a primary catalyst for many-body electronic instabilities. Because the phase space density $N(\epsilon_F)$ ($\epsilon_F$ is the Fermi energy) spikes exponentially or logarithmically at the exact critical filling fraction, the standard Fermi liquid description breaks down. Under the Bardeen-Cooper-Schrieffer (BCS) framework, the superconducting transition temperature $T_c$ scales non-linearly with the density of states \cite{polch}:
\begin{equation}
T_c \sim \hbar \omega_D \exp\left(-\frac{1}{V_{\text{eff}} N(\epsilon_F)}\right)\ ,
\end{equation}
where $\omega_D$ is the characteristic boson cutoff frequency and $V_{\text{eff}}$ is the effective attractive electron-electron coupling constant. As the filling fraction approaches the Van Hove singularity, $N(\epsilon_F) \to \infty$, collapsing the exponential suppression factor and maximizing $T_c$.
Here $T_c$ is assumed to be small so as to remain within the assumptions of \cite{zembatop}.

Concurrently, this local accumulation of phase space volume establishes an aggressive competition between distinct electronic phases. The enhanced density of states simultaneously satisfies the Stoner criterion for magnetic or structural ordering, driving collective instabilities such as:
{\it Spin density waves (SDW):} divergent magnetic susceptibility triggers real-space spatial modulations of the electron spin density.
{\it Charge density waves (CDW):} enhanced electron-phonon coupling freezes out periodic lattice distortions, opening a static structural gap.
Whether the system converges toward high-$T_c$ superconductivity or drops into a broken-symmetry density wave state depends heavily on the orientation of the lattice boundaries imposed by the underlying wallpaper group $\Gamma$.

{\it Non-orientable and exotic Fermi seas}:
beyond standard closed disk configurations, mapping free fermions to crystalline orbifolds $\mathbb{R}^2/\Gamma$ reveals non-orientable topological sectors within the framework of \cite{zembatop}. These states manifest in wallpaper groups containing glide reflection axes but lacking pure, orthogonal reflection lines {\it e.g.}, group $pg$).
\begin{figure}[h]
\centering
\textbf{Conceptual diagram: twisted boundary inversion} \\[0.1cm]
\small
\begin{tabular}{|lcl|}
\hline 
State vector $|\psi(k_x, k_y)\rangle$ & $\xrightarrow{\quad \text{glide operator } \hat{G} \quad}$ & $(-k_x, k_y + b/2) \otimes \mathcal{P}_{\text{twist}}$ \\ 
Orientation index $(+1)$ & $\xrightarrow{\quad \text{boundary crossing} \quad}$ & inverted orientation index $(-1)$ \\ \hline
\end{tabular} 
\end{figure}

When the fermion filling fraction forces the Fermi surface boundary to contact the edges of these glide-symmetric Brillouin zones, the physical space folds into global non-orientable structures:

{\it i) M\"obius strip Fermi boundaries:} under group $pg$, he programmed left/right coordinate assignment is reversed when moving along the Fermi boundary's perimeter $\ell$. Fermions must complete a double traversal of the apparent reciprocal zone loop ($4\pi$ rotation in momentum phase space) to map back onto their original quantum state vector.

{\it ii) Klein bottle topology:} generated by interlocking perpendicular glide operators ({\it e.g.}, group $pgg$). The Fermi sea behaves as an integrated manifold with no mathematically distinct interior volume or exterior boundary.

These structural distortions alter the reverse isoperimetric profile. Because the perimeter $\ell$ is topologically fixed around global non-orientable cycles, it cannot contract freely to minimize its boundary. Consequently, even at minimum filling configurations, exotic classes maintain an artificially inflated reverse isoperimetric baseline compared to their orientable counterparts, altering the geometric distribution of quantum states.

\subsection{Electrical conductivity as a function of the isoperimetric ratio}
The rigid displacement of the Fermi sea under a constant external electric field is a basic result of transport theory in conductors and semiconductors \cite{ashmer}. Usually, this displacement is derived from the semiclassical equations of motion of single charge carriers combined 
with the relaxation-time approximation. 
An alternative, non-perturbative formulation known as multidimensional bosonization or Haldane bosonization has emerged more recently \cite{haldane1994, netoneto}. This approach slices a high-dimensional Fermi surface into a collection of 1d channels. When an external electric field is applied, the steady-state rigid translation of the global Fermi surface can be reinterpreted as a collection of the charge flow due to the (1+1)d chiral anomalies 
occurring at the boundaries of each individual 1d slice \cite{else2021}. 

Consider a 2d free fermion gas with spin degeneracy $g_s = 2$. At zero temperature, the ground state forms a circular Fermi sea in momentum space defined by $k_x^2 + k_y^2 \le k_F^2$.
A uniform, constant external electric field $\mathbf{E} = E_x \hat{\mathbf{x}}$ along the $x$-direction is now applied. 
The Fermi sea may be partitioned as a continuous set of uncoupled parallel 1d slices along the $k_y$ axis. Each slice is parameterized by its transverse coordinate $k_y$, where $-k_F \le k_y \le k_F$.
\begin{equation}
k_{Fx}(k_y) = \sqrt{k_F^2 - k_y^2}\ .
\label{kxky}
\end{equation}
The boundaries of this segment intersect the circular Fermi surface at two distinct low-energy points: 
a {\it left-mover} located at $-k_{Fx}(k_y)$ and a {\it right-mover} located at $+k_{Fx}(k_y)$. For systems with large number of particles,
the fully occupied states deep within this 1d segment may be viewed as a 1d Dirac seas connecting these two counter-propagating edge modes.

When the electric field $E_x$ is turned on, every state accelerates according to the semiclassical equation of motion $\hbar {d\mathbf{k}}/{dt} = -e\mathbf{E}$ (for electrons with charge $q = -e$). Within an isolated 1d slice, this gives rise to 
a phenomenon known as {\it spectral flow} \cite{anomalies}. Electrons are continuously pumped out of the negative-energy vacuum at the left-moving Fermi point and are pushed across the bulk of the slice to emerge at the right-moving Fermi point.

For a single (1+1)d spinless chiral channel, Adler, Bell, and Jackiw showed that the conservation of individual chiral particle densities ($n_R$ and $n_L$) is broken by the electromagnetic field \cite{abj1, abj2}. The (1+1)d axial anomaly equation yielding the rate of change of the chiral particle density is:
\begin{equation}
\frac{\partial (n_R - n_L)}{\partial t} = \frac{e E_x}{2\pi \hbar}\ .
\end{equation}
Because every electron transferred from a left-moving state to a right-moving state alters the net balance, the total 1d particle production rate for a slice must account for the spin degeneracy factor $g_s$:
\begin{equation}
\frac{\partial N_{\text{1d}}}{\partial t} = g_s \cdot \frac{e E_x}{2\pi \hbar}\ .
\end{equation}

In a realistic clean non-relativistic electron gas, the continuous acceleration caused by the anomaly is balanced by scattering processes (such as electron-phonon 
interactions or impurity scattering). In the relaxation-time approximation, this dissipation is captured by a transport lifetime $\tau$. 
In this scenario, there is an effective steady-state time interval over which the anomaly operates before resetting, which is given by $\Delta t = \tau$. Therefore, the steady-state excess 1d particle density generated within that specific $k_y$-slice is:
\begin{equation}
\Delta N_{\text{1d}} = \frac{\partial N_{\text{1d}}}{\partial t} \tau = \frac{g_s e E_x \tau}{2\pi \hbar}\ .
\label{deltan}
\end{equation}

Each anomalous particle generated at the right-moving Fermi point carries a charge $-e$ and moves with the local group velocity component $v_x(k_y) = {\hbar k_{Fx}(k_y)}/{m}$,
which defines the effective mass $m$ of the charge carriers.
. The 1d electric current $I_{\text{1d}}$ carried by a single slice is given by the product of the charge, the excess density, and the velocity:
\begin{equation}
I_{\text{1d}}(k_y) = (-e) \Delta N_{\text{1d}}  v_x(k_y)\ .
\label{eqfori}
\end{equation}
Substituting Eqs. (\ref{kxky}) and (\ref{deltan}) into (\ref{eqfori}) yields:
\begin{equation}
I_{\text{1d}}(k_y) =  - \frac{g_s e^2 E_x \tau}{2\pi m} \sqrt{k_F^2 - k_y^2}\ .
\label{expfori}
\end{equation}

The global 2d macro-current density $J_x$ (which represents the current per unit cross-sectional length) is obtained by integration of these individual 1d slice currents over all possible values of $k_y$. Since the transverse spatial density of states per unit momentum is $1/(2\pi)$, the integration is 
written as:
\begin{equation}
J_x = \int_{-k_F}^{k_F} \frac{dk_y}{2\pi} I_{\text{1d}}(k_y)\ .
\end{equation}
Substition of (\ref{expfori}) leads to an integral that represents the area of a semi-circle of radius 
$k_F$,{\it i.e.}, $\pi k_F^2/2$. Thus, the current density becomes:
\begin{equation}
J_x = - \frac{g_s e^2 E_x \tau}{4\pi^2 m} \left( \frac{\pi k_F^2}{2} \right) = - \frac{g_s e^2 k_F^2 \tau}{8\pi m} E_x\ .
\label{condfree}
\end{equation}
This expression can be written in terms of standard thermodynamic variables, noting that the Fermi wavevector $k_F$ is related to the 
total 2d spatial carrier density $n$. Given the spin degeneracy $g_s$:
\begin{equation}
n = g_s \cdot \frac{\cal{A}}{(2\pi)^2} = g_s  \cdot \frac{\pi k_F^2}{4\pi^2} = g_s \frac{k_F^2}{4\pi} \implies k_F^2 = 
\frac{4\pi n}{g_s}\ .
\end{equation}
Substituting $k_F^2 = 4\pi n/g_s $ back into our equation for the current density magnitude $|J_x|$ yields:
\begin{equation}
|J_x| = \frac{g_s e^2 (4\pi n) \tau}{g_s 4\pi m} E_x = \frac{n e^2 \tau}{m} E_x\ .
\end{equation}
Invoking Ohm's Law ($J_x = \sigma_{xx} E_x$), the 2d electrical conductivity is found to be:
\begin{equation}
\sigma_{xx} = \frac{n e^2 \tau}{m}\ ,
\label{drude}
\end{equation}
which is of the Drude-Sommerfeld form.

The above reasoning can be extended to the case of a gas of 2d free charges particles moving on a lattice background.
A formal link between electric conductivity and $\mathcal{I}$, matching the contents of Table \ref{tab:zemba_classes} may be established 
either by an extension of Eq. (\ref{condfree}) or, independently,
via the semi-classical Boltzmann transport equation (or the $T=0$ limit of the Kubo formula) for the longitudinal conductivity $\sigma_{xx}$ in a 2d sheet.
Note that, from Eq. (\ref{condfree}), $\sigma_{xx}$ may be written in terms of the length of the Fermi sea boundary:
\begin{equation}
    \sigma_{xx} = \frac{g_s e^2 \tau}{8\pi^2 \hbar} \cdot \frac{1}{2} v_F {\ell}\ .
\end{equation}
The geometric generalization of it for the orbifold Fermi seas $\cal O$ that preserves the same structure is given by: 
\begin{equation}
    \sigma_{xx} = \frac{g_s e^2 \tau}{8\pi^2 \hbar} \oint_{\partial\cal O} \frac{v_x^2}{|\mathbf{v}|} dl\ ,
\label{sigmaxx}
\end{equation}
where $dl$ is the line element along the Fermi boundary $\partial\cal O$, $\mathbf{v} = \frac{1}{\hbar} \nabla_\mathbf{k} \epsilon_\mathbf{k}$ is the group velocity vector, $v_x$ is its projection along the applied electric field, 
and $\tau$ is the {\it average relaxation time} over the Fermi boundary. 
In this sense, the formula for $\sigma_{xx}$ is a {\it mean field theory} approximation, adopted to keep the discussion 
in simple terms aiming to isolate the dependence of it on $\cal{I}$.  
For simplicity, we assume that the Fermi boundary has one simply-connected component. The extension to other cases, {\it e.g.}, the annulus, is straightforward.
In order to parameterize the structural line integral, it is convenient to define an effective (average) Fermi velocity $v_F$ and a directional orientation average factor $\gamma_x$:
\begin{equation}
    \oint_{\partial\cal O} \frac{v_x^2}{|\mathbf{v}|} dl = \gamma_x v_F {\ell}\ ,
\end{equation}
where ${\ell} = \oint_{\partial\cal O}dl$.
Substituting this in (\ref{sigmaxx}) implies a transport expression explicitly linear in the boundary length:
\begin{equation}
    \sigma_{xx} = \left( \frac{g_s e^2 \tau \gamma_x v_F}{8\pi^2 \hbar} \right) \cdot {\ell}\ .
\end{equation}
which yields:
\begin{equation}
    \sigma_{xx} = \left[ \frac{g_s e^2 \tau \gamma_x v_F {\cal A}}{2\pi \hbar {\ell}} \right] \cdot \frac{1}{\sqrt\mathcal{I}}\ .
\end{equation}
Invoking Luttinger's theorem for a 2d system, the enclosed momentum area encodes the carrier density $n$ via $n = g_s {\cal A}/(2\pi)^2$, one obtains:
\begin{equation}\label{eq:unification}
    \sigma_{xx} = \left( \frac{2\pi e^2 \tau \gamma_x v_F n}{\hbar {\ell}} \right) \cdot \frac{1}{\sqrt\mathcal{I}}\ .
\end{equation}
Upon defining the effective mass of the charge carriers by $m = \hbar {\ell}/(4\pi \gamma_{x} v_F)$,
and noting that $\ell \propto k_F$, a scaling relation of the conductivity with $\cal{I}$ is found:
\begin{equation}\label{eq:sigscaling}
    \sigma_{xx} ({\cal I})= \sigma_{xx} (1)\cdot \frac{1}{\sqrt\mathcal{I}}\ ,
\end{equation}
where $\sigma_{xx} (1)$ is defined by Eq. (\ref{drude}). This result exposes the scaling of the electric conductivity with $\cal{I}$ in 2d. Note that the effective mass $m$ defined after (\ref{eq:unification}) coincides with that 
after (\ref{deltan}) for the isotropic case (in which $\gamma_{x}=1/2$).

To summarize, the macroscopic transport properties of the conducting classes may be characterized by a 
topological-geometric signature given by $1/\sqrt \mathcal{I}$. This relates the {\it rigidity} or {\it compactness} of the ground-state fermion array to the large-scale material behavior, which is, in turn, measured by $\mathcal{I}$. 
For high compactness ($\mathcal{I} \to 1$) the Fermi sea resembles a perfect disk. The ground state minimizes kinetic energy for a given density and the arrangement is maximally rigid against directional distortions, representing isotropic Drude-like metallic transport.
For low compactness ($\mathcal{I} \ll 1$) crystalline potentials or orbifold constraints highly deform, stretch, or corrugate the Fermi boundary. While the area ${\cal A}$ (carrier density) remains constant, the boundary perimeter $\ell$ grows ($\mathcal{I} \to 0$). The Fermi sea loses circular compactness, triggering highly anisotropic, directional, or enhanced conduction channels.

\subsubsection{Orbifold corrections to the relaxation time}
In the flat orbifold classification boundaries are non-smooth and contain a discrete number of singular conical points and reflection corners ($N_c$). These sharp geometric distortions act as topologically protected, intrinsic scattering centers. 
The quantum mechanical relaxation time $\tau$ must be modified to account for this structural drag. A proposed phenomenological expression is:
\begin{equation}
    \tau = \frac{\tau_0}{1 + \alpha N_c}\ ,
\end{equation}
where $\tau_0$ is the bulk impurity relaxation time and $\alpha$ is a coupling coefficient reflecting the scattering cross-section of an orbifold corner. 
The actual presence of $\tau_0$ is attributed to residual background disorder/phonons decoupled from the orbifold geometry, that is, enlarging 
the scope of the discussion of this work, a reasonable assumption relying on the topological nature of the classification \cite{zembatop} and, therefore, on the universality of its classes. 
Substituting this into Eq.\,(\ref{eq:unification}) yields a final unified expression for transport of topological conducting classes:
\begin{equation}
    \sigma_{xx}^{\text{orbifold}} = \left( \frac{2\pi e^2 \gamma_x v_F n}{\ell} \cdot \frac{\tau_0}{(1 + \alpha N_c)} \right) \cdot \frac{1}{\sqrt\mathcal{I}}\ .
\end{equation}

\subsection{Conclusions}

Flat orbifold classification of Fermi seas unifies microscopic discrete lattice symmetry with mesoscopic coordinate space density modulations and 
macroscopic transport optimization, superseding standard EFT limitations regarding short-scale symmetry resilience \cite{boze}. 
By coupling geometric measure theory (isoperimetric packing and vertex/boundary weight accounting) with chiral anomaly/relaxation transport descriptions, 
the framework establishes a geometric-topological phase descriptor, the isoperimetric ratio $\cal{I}$, laying a basis for experimental exploration of Fermi sea geometries at ultra-low temperatures \cite{fermitop2026}.
The classification and investigation of the electric characteristics of Fermi seas in 3d might benefit from the application of the isoperimetric ratio due to its simplicity.

%

\end{document}